\documentclass[singlecolumn,epjc3]{svjour3}

\smartqed  % flush right qed marks, e.g. at end of proof

\RequirePackage{graphicx}

\usepackage{amsmath,amssymb}

\newcommand*\DAlambert{\mathop{}\!\mathbin\Box}

\journalname{Eur. Phys. J. C}

\begin{document}

\title{Compact stars in $f(R,\mathcal{T})$ gravity}

\author{Amit Das\thanksref{e1,addr1}
\and Farook  Rahaman\thanksref{e2,addr2} \and B.K.
Guha\thanksref{e3,addr3} \and Saibal Ray\thanksref{e4,addr4}. }

\thankstext{e1}{e-mail: amdphy@gmail.com}
\thankstext{e2}{e-mail: rahaman@associates.iucaa.in}
\thankstext{e3}{e-mail: bkguhaphys@gmail.com}
\thankstext{e4}{e-mail: saibal@associates.iucaa.in}

\institute{Department of Physics, Indian Institute of Engineering
Science and Technology, Shibpur, Howrah 711103, West Bengal,
India\label{addr1}
              \and
              Department of Mathematics, Jadavpur University, Kolkata
700032, West Bengal, India\label{addr2}
              \and
              Department of Physics, Indian
Institute of Engineering Science and Technology, Shibpur, Howrah
711103, West Bengal, India\label{addr3}
              \and
               Department of Physics, Government College of Engineering and Ceramic Technology, Kolkata
700010, West Bengal, India\label{addr4} }

\date{Received: date / Accepted: date}

\maketitle

\begin{abstract}
In the present paper we generate a set of solutions describing the
interior of a compact star under $f(R,\mathcal{T})$ theory of
gravity which admits conformal motion.  An extension of general
relativity, the $f(R,\mathcal{T})$ gravity is associated to Ricci
scalar $R$ and the trace of the energy-momentum tensor
$\mathcal{T}$. To handle the Einstein field equations in the form
of differential equations of second order, first of all we adopt
the Lie algebra with conformal Killing vectors (CKV) which enable
one to get solvable form of such equations and second we consider
the equation of state (EOS) $p=\omega\rho$ with $0<\omega<1$ for
the fluid distribution consisting of normal matter, $\omega$ being
the EOS parameter. We therefore analytically explore several
physical aspects of the model to represent behaviour of the
compact stars  such as - energy conditions, TOV equation,
stability of the system, Buchdahl condition, compactness and
redshift. It is checked that the physical validity and the
acceptability of the  present model within the specified
observational constraint in connection to a dozen of the compact
star candidates are quite satisfactory.
\end{abstract}

\section{Introduction}
Though Einstein's general theory of relativity has always proved
to be very fruitful for uncovering so many hidden mysteries of
Nature, yet the evidence of late-time acceleration of the Universe
and the possible existence of dark matter has imposed a
fundamental theoretical challenge to this
theory~\cite{Ri1998,Perl1999,Bern2000,Hanany2000,Peebles2003,Paddy2003,clifton2012}.
As a result, several modified theories on gravitation have been
proposed from time to time. Among all these theories, a few of
them, namely $f(R)$ gravity, $f(T)$ gravity and $f(R,\mathcal{T})$
gravity, have received more attention than any other. In all these
theories instead of changing the source side of the Einstein field
equations, the geometrical part has been changed by taking a
generalized functional form of the argument to address galactic,
extra-galactic, and cosmic dynamics. Cosmological models based
upon modified gravity theories reveal that excellent agreement
between theory and observation can be
obtained~\cite{hwang2001,bahcall1999,demianski2006,singh2015}.

In $f(R)$ gravity theory the gravitational part in the standard
Einstein-Hilbert action is replaced by an arbitrary generalized
function of the Ricci scalar $R$ whereas in $f(T)$ gravity theory
the same is replaced by an arbitrary analytic function of torsion
scalar $T$. The $f(T)$ theory of gravity is more controllable than
$f(R)$ theory of gravity because the field equations in the former
turns out to be the differential equations of second order whereas
in the later the field equations in the form of differential
equations are, in general, of fourth order, which is difficult to
handle~\cite{Boehmer2011}. Many applications of $f(T)$ gravity in
cosmology, theoretical presentation as well as observational
verification, can be found in
Refs.~\cite{Wu2010a,Tsyba2011,Dent2011,Chen2011,Bengochea2011,Wu2010b,Yang2011,Zhang2011,Li2011,Wu2011,Bamba2011,Krssak2015,Nassur2016a,Nassur2016b,Cai2015,Bamba2016}.
On the other hand, many astrophysical applications of $f(T)$
theory of gravity can be observed in
Refs.~\cite{Boehmer2011,Deliduman2011,Wang2011,Daouda2011,Abbas2015a,Abbas2015b}.
Following the result of B\"{o}hmer et al.~\cite{Boehmer2011} in
our previous work~\cite{Das2015} we successfully described the
interior of a relativistic star along with the existence of a
conformal Killing vector field within this $f(T)$ gravity
providing a set of exact solutions. In connection to $f(R)$
gravity we observe that there are also several applications with
various aspects on the theory available in the
literature~\cite{Carroll2004,Capozziello2006,Nojiri2006}. A
special and notable application includes about the late-time
acceleration of the Universe which has been explained using $f(R)$
gravity by Carroll et al.~\cite{Carroll2004}. For further reviews
on $f(R)$ gravity model one can check
Refs.~\cite{Nojiri2011,Soti2010,Lobo2008,Capozziello2010,Capozziello2011}.

However, the purpose of the present paper is to consider another
extension of general relativity, the $f(R,\mathcal{T})$  modified
theory of gravity~\cite{harko2011} where the gravitational
Lagrangian of the standard Einstein-Hilbert action is defined by
an arbitrary function of the Ricci scalar $R$ and the trace of the
energy-momentum tensor $\mathcal{T}$. It has been argued that such
a dependence on $\mathcal{T}$ may come from the presence of
imperfect fluid or quantum effects. Many cosmological applications
based on the $f(R,\mathcal{T})$ gravity can be found in
~\cite{moraes2014b,moraes2015a,moraes2015b,singh2014,rudra2015,baffou2015,shabani2013,shabani2014,sharif2014b,ram2013,reddy2013b,kumar2015,shamir2015,Fayaz2016}.

Though one can find several applications to astrophysical level
based on this theory, yet among those it is worth to mention
Refs.~\cite{sharif2014,noureen2015,noureen2015b,noureen2015c,zubair2015a,zubair2015b,Ahmed2015,Moraes2015}.
Sharif et al.~\cite{sharif2014} have discussed the stability of
collapsing spherical body of an isotropic fluid distribution
considering the non-static spherically symmetric line element. On
the other hand, a perturbation scheme has been used to find the
collapse equation and the condition on the adiabatic index has
been developed for Newtonian and post-Newtonian eras for
addressing instability problem by Noureen et
al.~\cite{noureen2015}. Further, Noureen et
al.~\cite{noureen2015b} have developed the range of instability
under the $f(R,\mathcal{T})$ theory for an anisotropic background
constrained by zero expansion. The evolution of a spherical star
by applying a perturbation scheme on the $f(R,\mathcal{T})$ field
equations has been explored by Noureen et al.~\cite{noureen2015c},
while in the work~\cite{zubair2015a} the dynamical analysis for
gravitating sources along with axial symmetry has been discussed.
Zubair et al.~\cite{zubair2015b} investigated the possible
formation of compact stars in $f(R,\mathcal{T})$ theory of gravity
using analytic solution of the Krori and Barua metric to the
spherically symmetric anisotropic star. The effects of
$f(R,\mathcal{T})$ gravity on gravitational lensing has been
discussed by Ahmed et al.~\cite{Ahmed2015}. Moraes et
al.~\cite{Moraes2015} have investigated the spherical equilibrium
configuration of polytropic and strange stars under
$f(R,\mathcal{T})$ theory of gravity.

Using the technique of CKV one can search for the inheritance
symmetry which provides a natural relationship between geometry
and matter through the Einstein field equation. Several works
performed by using this technique of conformal motion to the
astrophysical field can be found in the following
Refs.~\cite{Das2015,Ray2008,Rahaman2010a,Rahaman2010b,Usmani2011,Bhar2014,Rahaman2014,Rahaman2015b,Rahaman2015c}.
Interior solutions admitting conformal motions also had been
studied extensively by Herrera et al.
~\cite{Herrera1984,Herrera1985a,Herrera1985b,Herrera1985c}. An
exact solution describing the interior of a charged quark star had
been explored admitting a one-parameter group of conformal motions
by Mak and Harko~\cite{Harko2004}.

In the present work we shall seek the interior solutions of the
Einstein field equations under the $f(R,\mathcal{T})$ theory of
gravity along with conformal Killing vectors. Therefore, our main
aim in the present work is to construct a set of stellar solutions
under $f(R,\mathcal{T})$ theory of gravity by assuming the
existence of Conformal Killing Vectors (CKVs). The outline of our
investigation is as follows: in Sect. 2 we provide the basic
mathematical formalism of $f(R,\mathcal{T})$ theory whereas the
CKVs have been formulated in Sect. 3. In Sect. 4 we provide the
field equations under $f(R,\mathcal{T})$ gravity along with their
solutions using the technique of CKV, whereas in Sect. 5 the
exterior Schwarzschild solution and matching conditions are
provided. In Sect. 6 we discuss some physical features of the
model such as energy conditions and the equilibrium condition by
using Tolman-Oppenheimer-Volkoff (TOV) equation, the stability
issue, the mass-radius relation, compactness, and surface
redshift. A comparative study for the physical validity of the
model is performed in Sect. 7. Lastly, in Sect. 8 we make some
concluding remarks.

\section{Basic mathematical formalism of the $f(R,\mathcal{T})$ Theory}

The action of the $f(R,\mathcal{T})$ theory~\cite{harko2011} is
taken as
\begin{equation}\label{eq1}
S=\frac{1}{16\pi}\int d^{4}xf(R,\mathcal{T})\sqrt{-g}+\int
d^{4}x\mathcal{L}_m\sqrt{-g},
\end{equation}
where $f(R,\mathcal{T})$ is an arbitrary function of the Ricci
scalar $R$ and the trace of the energy-momentum tensor
$\mathcal{T}$ and $\mathcal{L}_m$ being the Lagrangian for matter.
Also $g$ is the determinant of the metric $g_{\mu\nu}$. Here we
assume the geometrical units $G=c=1$.

If one varies the action (\ref{eq1}) with respect to the metric
$g_{\mu\nu}$, one can get the following field equations of
$f(R,\mathcal{T})$ gravity:
\begin{eqnarray}\label{eq2}
f_R (R,\mathcal{T}) R_{\mu\nu} - \frac{1}{2} f(R,\mathcal{T}) g_{\mu\nu} + (g_{\mu\nu}\DAlambert - \nabla_{\mu} \nabla_{\nu}) f_R (R,\mathcal{T})\nonumber \\
= 8\pi T_{\mu\nu} - f_\mathcal{T}(R,\mathcal{T}) T_{\mu\nu} -
f_\mathcal{T}(R,\mathcal{T})\Theta_{\mu\nu},
\end{eqnarray}
where $f_R (R,\mathcal{T})= \partial f(R,\mathcal{T})/\partial R$,
$f_\mathcal{T}(R,\mathcal{T})=\partial f(R,\mathcal{T})/\partial
\mathcal{T}$, $\DAlambert \equiv
\partial_{\mu}(\sqrt{-g} g^{\mu\nu} \partial_{\nu})/\sqrt{-g}$,
$R_{\mu\nu}$ is the Ricci tensor, $\nabla_\mu$ provides the
covariant derivative with respect to the symmetric connection
associated to $g_{\mu\nu}$, $\Theta_{\mu\nu}=
g^{\alpha\beta}\delta T_{\alpha\beta}/\delta g^{\mu\nu}$ and the
stress-energy tensor can be defined as
$T_{\mu\nu}=g_{\mu\nu}\mathcal{L}_m-2\partial\mathcal{L}_m/\partial
g^{\mu\nu}$.

The covariant divergence of (\ref{eq2}) reads as~\cite{barrientos2014}
\begin{eqnarray}\label{eq3}
\hspace{-0.5cm}\nabla^{\mu}T_{\mu\nu}&=&\frac{f_\mathcal{T}(R,\mathcal{T})}{8\pi -f_\mathcal{T}(R,\mathcal{T})}[(T_{\mu\nu}+\Theta_{\mu\nu})\nabla^{\mu}\ln f_\mathcal{T}(R,\mathcal{T})\nonumber \\
&&+\nabla^{\mu}\Theta_{\mu\nu}-(1/2)g_{\mu\nu}\nabla^{\mu}\mathcal{T}].
\end{eqnarray}

Equation (\ref{eq3}) at once shows that the energy-momentum tensor
is not conserved for the $f(R,\mathcal{T})$ theory of gravity
unlike in the case of general relativity.

In this paper we assume the energy-momentum tensor to be that of a
perfect fluid, i.e.
\begin{equation}\label{eq4}
T_{\mu\nu}=(\rho+p)u_\mu u_\nu-pg_{\mu\nu},
\end{equation}

with $u^{\mu}u_{\mu} = 1$ and $u^\mu\nabla_\nu u_\mu=0$. Also with
these conditions we have $\mathcal{L}_m=-p$ and
$\Theta_{\mu\nu}=-2T_{\mu\nu}-pg_{\mu\nu}$.

As proposed by Harko et al.~\cite{harko2011}, we have taken the
functional form of $f(R,\mathcal{T})$ as $f(R,\mathcal{T})=R+2\chi
\mathcal{T}$, where $\chi$ is a constant. We note that this form
has been extensively used to obtain many cosmological solutions in
$f(R,\mathcal{T})$
gravity~\cite{singh2015,moraes2014b,moraes2015a,moraes2015b,reddy2013b,kumar2015,shamir2015}.
After substituting the above form of $f(R,\mathcal{T})$ in
(\ref{eq2}), one can get~\cite{moraes2014b,moraes2015a}
\begin{equation}\label{eq5}
G_{\mu\nu}=8\pi T_{\mu\nu}+\chi \mathcal{T}
g_{\mu\nu}+2\chi(T_{\mu\nu}+pg_{\mu\nu}),
\end{equation}
where $G_{\mu\nu}$ is the Einstein tensor.

One can easily get back to the general relativistic result just by
setting $\chi=0$ in the above Eq. (\ref{eq5}) . Moreover, for
$f(R,\mathcal{T})=R+2\chi \mathcal{T}$, Eq. (\ref{eq3}) reads
\begin{equation}\label{eq6}
(8\pi+2\chi)\nabla^{\mu}T_{\mu\nu}=-2\chi\left[\nabla^{\mu}(pg_{\mu\nu})+\frac{1}{2}g_{\mu\nu}\nabla^{\mu}\mathcal{T}\right].
\end{equation}
Again substituting $\chi=0$ in Eq. (\ref{eq6}) one can easily
verify that the energy-momentum tensor is conserved as in the case
of general relativity.

\section{The Conformal Killing Vector (CKV)}
To search a natural relationship between geometry and matter
through Einstein's general relativity one can use symmetries.
Symmetries that arise either from a geometrical viewpoint or
physical relevant quantities are known as collineations. The
greatest advantageous collineations is the conformal Killing
vectors (CKV). Those vectors also provide a deeper insight into
the spacetime geometry. From a mathematical viewpoint, conformal
motions or conformal Killing vectors (CKV) are motions along which
the metric tensor of a spacetime remains invariant up to a scale
factor. Moreover, the advantage of using the CKV is that it
facilitates the generation of exact solutions to the field
equations. Also using the technique of CKV one can easily reduce
the highly nonlinear partial differential equations of Einstein's
gravity to ordinary differential equations.

The CKV is defined as
\begin{equation}
L_{\xi} g_{ij} = \xi_{i;j}+ \xi_{j;i} = \psi g_{ij},\label{eq7}
\end{equation}
where $L$ is the Lie derivative operator, which describes the
interior gravitational field of a stellar configuration with
respect to the vector field $\xi$ and $\psi$ is the conformal
factor. One can note that the vector $\xi$ generates the conformal
symmetry and the metric $g$ is conformally mapped onto itself
along $\xi$. However, B\"{o}hmer et
al.~\cite{Bohmer2007,Bohmer2008} argued that neither $\xi$ nor
$\psi$ need to be static even though a static metric is
considered. We also note that (i) if $\psi=0$ then Eq. (7) gives
the Killing vector, (ii) if $\psi=$ constant it gives homothetic
vector, and (iii) if $\psi=\psi(\textbf{x},t)$ then it yields
conformal vectors. Moreover, for $\psi=0$ the underlying spacetime
becomes asymptotically flat which further implies that the Weyl
tensor will also vanish. All these properties reflect that CKV has
an intrinsic property to providing deeper insight of the
underlying spacetime geometry.

Under the above background, let us therefore consider that our
static spherically symmetric spacetime admits an one parameter
group of conformal motion. In this case the metric can be opted as
\begin{equation}
ds^2=-e^{\nu(r)}dt^2+e^{\lambda(r)}dr^2+r^2(d\theta^2+\sin^2\theta
d\phi^2),\label{eq8}
\end{equation}
which is conformally mapped onto itself along $\xi$. Here $\nu$ and $\lambda$
are metric potentials and functions of the radial coordinate $r$ only.

Here Eq. (\ref{eq7}) implies that
\begin{equation}
L_\xi g_{ik} =\xi_{i;k}+ \xi_{k;i} = \psi g_{ik},\label{eq9}
\end{equation}
with $\xi_i = g_{ik}\xi^k$.

From Eqs. (\ref{eq8}) and (\ref{eq9}), one can find the following
expressions~\cite{Herrera1985a,Herrera1985b,Herrera1985c,Harko2004}:
\begin{eqnarray}
&\xi^1 \nu^\prime =\psi,\nonumber \\
&\xi^4  = {\rm constant},\nonumber \\
&\xi^1  = \frac{\psi r}{2},\nonumber \\
&\xi^1 \lambda^\prime + 2 \xi^1 _{,1} =\psi, \nonumber
\end{eqnarray}
where $1$ and $4$ stand for the spatial and temporal coordinates
$r$ and $t$, respectively.

From the above set of equations one can get
\begin{eqnarray}
e^\nu  &=& C_2^2 r^2, \label{eq10}\\ e^\lambda  &=&
\left[\frac {C_3} {\psi}\right]^2,  \label{eq11} \\ \xi^i &=& C_1
\delta_4^i + \left[\frac{\psi r}{2}\right]\delta_1^i, \label{eq12}
\end{eqnarray}
where $C_1$, $C_2$, and $C_3$ all are integration constants.

\section{The field equations and their solutions in $f(R,\mathcal{T})$ gravity}

For the spherically symmetric metric (\ref{eq8}) one can find the
non-zero components of the Einstein tensors as
\begin{equation}
G_0^{0}=\frac{e^{-\lambda}}{r^{2}}(-1+e^{\lambda}+\lambda'
r),\label{eq13}
\end{equation}

\begin{equation}
G_1^{1}=\frac{e^{-\lambda}}{r^{2}}(-1+e^{\lambda}-\nu'
r),\label{eq14}
\end{equation}

\begin{equation}
G_2^{2}=G_3^{3}=\frac{e^{-\lambda}}{4r}[2(\lambda'-\nu')-(2\nu''+\nu'^{2}-\nu'\lambda')r],\label{eq15}
\end{equation}
where primes stand for derivations with respect to the radial coordinate $r$.

Substituting Eqs. (4), (\ref{eq13}), and (\ref{eq14}) in Eq.
(\ref{eq5}) one can get
\begin{equation}
-1+e^{\lambda}+\lambda'r=\Pi(r)[8\pi\rho+\chi(3\rho-p)].\label{eq16}
\end{equation}

\begin{equation}
-1+e^{\lambda}-\nu'r=\Pi(r)[-8\pi p+\chi(\rho-3p)], \label{eq17}
\end{equation}
with $\Pi(r)\equiv r^{2}/e^{-\lambda}$.

Now using Eqs. (\ref{eq10}), (\ref{eq11}), (\ref{eq16}), and
(\ref{eq17}) one can obtain
\begin{equation}
-\frac{2\psi\psi'}{rC_3^2}-\frac{\psi^2}{r^2C_3^2}+\frac{1}{r^2}
=[8\pi\rho+\chi(3\rho-p)].\label{eq18}
\end{equation}

\begin{equation}
-\frac{3\psi^2}{r^2C_3^2}+\frac{1}{r^2}=[-8\pi
p+\chi(\rho-3p)].\label{eq19}
\end{equation}

To solve the Eqs. (\ref{eq18}) and (\ref{eq19}) let us assume the
equation of state of fluid distribution consisting of normal
matter as
\begin{equation}
p = \omega \rho,\label{eq20}
\end{equation}
where $\omega$ is the equation of state parameter, with
$0<\omega<1$.

Inserting Eq. (\ref{eq20}) in Eqs. (\ref{eq18}) and (\ref{eq19})
we, respectively, get
\begin{equation}
\rho= -\frac{1}{\varepsilon
C_3^2r}\left[2\psi\psi'+\frac{1}{r}(\psi^2-C_3^2)\right],
\label{eq21}
\end{equation}
and
\begin{equation}
\rho=-\frac{1}{\alpha r^2}\left[\frac{3\psi^2}{C_3^2}- 1
\right],
\label{eq22}
\end{equation}
where $\varepsilon$ and $\alpha$ are given by
$\varepsilon=\left[8\pi+\chi(3-\omega)\right],~\alpha=\left[-8\pi\omega+\chi(1-3\omega)\right]$,
respectively.

Now equating the above two expressions of the density $\rho$  we
have found the following differential equation in $\psi$:
\begin{equation}
-\left(\frac{2}{C_3^2}\right)r\psi\psi'-\left(\frac{\beta}{C_3^2}\right)\psi^2+\sigma=0.\label{eq23}
\end{equation}

Solving Eq. (\ref{eq23}) one can obtain the following solution
set:
\begin{equation}
\psi^2=\left[kC_3^2r^{-\beta}+
\frac{C_3^2\sigma}{\beta}\right],\label{eq24}
\end{equation}

\begin{equation}
\rho=\left[-3kr^{-\beta}-\frac{3\sigma}{\beta}+1\right]\times
\left(\frac{r^{-2}}{\alpha}\right),\label{eq25}
\end{equation}

\begin{equation}
p=\omega\left[-3kr^{-\beta}-\frac{3\sigma}{\beta}+1\right]\times
\left(\frac{r^{-2}}{\alpha}\right),\label{eq26}
\end{equation}
where $\beta$ and $\sigma$ are given by
$\beta=\left[\frac{8\pi\omega+8\chi+24\pi}{\omega(8\pi+3\chi)-\chi}\right],~
\sigma=\left[\frac{\omega(8\pi+2\chi)+2\chi+8\pi}{\omega(8\pi+3\chi)-\chi}\right]$,
respectively, and $k$ is an integration constant.

\section{The exterior Schwarzschild solution and matching conditions}

The well-known static exterior Schwarzschild solution is given by
\begin{equation}
ds^2=-\left(1-\frac{2M}{r}\right)dt^2+\left(1-\frac{2M}{r}\right)^{-1}dr^2+r^2\left(d\theta^2+sin^2\theta
d\phi^2\right).\label{eq27}
\end{equation}

For the continuity of the metric namely $g_{tt}$ and $g_{rr}$
across the boundary i.e. $r=a$ we have the following equations:
\begin{equation}
C_2^2=\frac{1}{a^2}\left(1-\frac{2M}{a}\right),\label{eq28}
\end{equation}

\begin{equation}
\left(ka^{-\beta}+\frac{\sigma}{\beta}\right)=
\left(1-\frac{2M}{a}\right).\label{eq29}
\end{equation}

Also at the boundary (i.e. $r=a$) the pressure $p=0$. Hence we
have
\begin{equation}
\left(-3ka^{-\beta}-\frac{3\sigma}{\beta}+1\right)=0.\label{eq30}
\end{equation}

The constant $C_2$ can be determined from Eq. (\ref{eq28}). But
Eqs. (29) and (30) are not independent equations. Thus, we have
only one independent equation with two unknowns, namely the
integration constant $k$ and $\chi$. So, in principle, these
equations are redundant to solve for $k$ and $\chi$.

\section{Physical features of the model under $f(R,\mathcal{T})$ gravity}

\subsection{Energy conditions}

To check whether all the energy conditions are satisfied or not
for our model under $f(R,\mathcal{T})$ gravity we should consider
the following inequalities:
\[
(i)~NEC: \rho+p_r\geq 0,~\rho+p_t\geq 0,
\]
\[
(ii)~WEC: \rho+p_r\geq 0,~\rho\geq 0,~\rho+p_t\geq 0,
\]
\[
(iii)~SEC: \rho+p_r\geq 0,~\rho+p_r+2p_t\geq 0.
\]
Here for our model of an isotropic fluid distribution (i.e. $p_r =
p_t = p$) we see from Fig. 2 that all the solutions are physically
valid. However, the behaviour of density and pressure is shown in
Fig. 1.

\begin{figure}[h]
\centering
\includegraphics[width=0.4\textwidth]{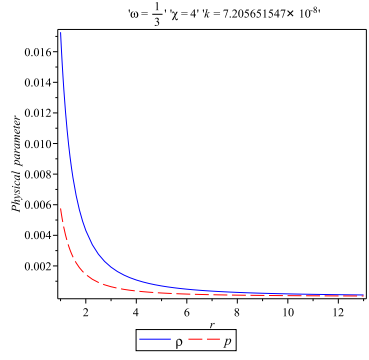}
\caption{Variations of density $\rho$ ($km^{-2}$) and pressure $p$
($km^{-2}$) is shown with respect to the radial coordinate $r$
($km$)}
\end{figure}

\begin{figure}[h]
\centering
\includegraphics[width=0.4\textwidth]{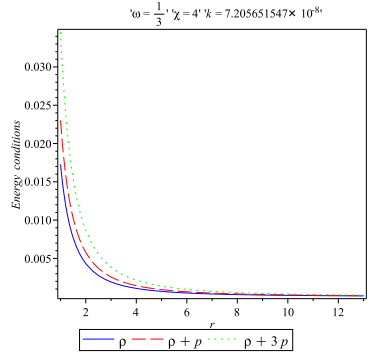}
\caption{Variations of $\rho$ ($km^{-2}$), $\rho + p$ ($km^{-2}$),
and $\rho+3p$ ($km^{-2}$) are shown with respect to the radial
coordinate $r$ ($km$)}
\end{figure}

\subsection{TOV equation}
From the equation for the non-conservation of the energy-momentum
tensor in $f(R,\mathcal{T})$ theory (\ref{eq6}) one can obtain the
generalized Tolman-Oppenheimer-Volkoff (TOV)
equation~\cite{Moraes2015} for an isotropic fluid distribution
(i.e. $p_r = p_t = p$) as
\begin{equation}
-\frac{\nu'}{2}(\rho+p)-\frac{dp}{dr}+\frac{\chi}{8\pi+2\chi}(p'-\rho')=0,\label{eq31}
\end{equation}

If one puts $\chi=0$ then one can get the usual form of TOV
equation in the case of general relativity. The above TOV equation
describes the equilibrium of the stellar configuration under the
joint action of three forces, viz. the gravitational force
($F_g$), the hydrostatic force ($F_h$), and the additional force
($F_\varkappa$) due to the modification of the gravitational
Lagrangian of the standard Einstein-Hilbert action. So for
equilibrium condition one can eventually write it in the following
form:
\begin{equation}
F_g+F_h+F_\varkappa=0,\label{eq32}
\end{equation}
where
\[F_g=-\frac{\nu'}{2}(\rho+p),\]
\[F_h=-\frac{dp}{dr},\]
\[F_\varkappa=\frac{\chi}{8\pi+2\chi}(p'-\rho')\].

In the present conformally symmetric model of an isotropic fluid
distribution with the EOS $ p = \omega \rho$ the TOV equation
(\ref{eq31}) can be written as

\begin{equation}
-\frac{\nu'}{2}(\rho+p)-\frac{dp}{dr}+\frac{\chi}{8\pi+2\chi}(\omega-1)\rho'=0.\label{eq33}
\end{equation}

From Fig. 3 we notice that the static equilibrium has been
attained under the mutual action of the three forces $F_g$, $F_h$
and $F_\varkappa$. Also it is observed from the figure that $F_g$
and $F_\varkappa$ are essentially of the same nature -
quantitatively as well as qualitatively.

\begin{figure}[h]
\centering
\includegraphics[width=0.4\textwidth]{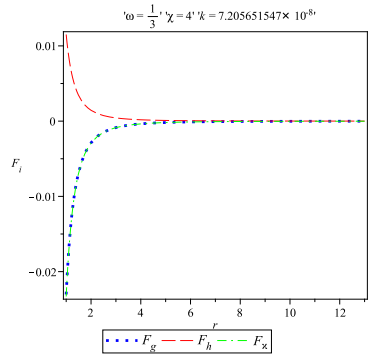}
\caption{The three different forces, viz. the gravitational force
$(F_g)$, the hydrostatic force $(F_h)$, and the additional force
$(F_\varkappa)$ are plotted against $r$ ($km$)}
\end{figure}

\subsection{Stability}

\subsubsection{Sound Speed}
According to Herrera~\cite{herrera1992} for a physically
acceptable model the square of the sound speed, i.e. $
v_s^{2}=\frac{dp}{d\rho}$, within the matter distribution should
be in the limit [0,1]. In our model of an isotropic matter
distribution we see that $v_s^{2}=\frac{dp}{d\rho}=\omega=1/3~<1$.
Hence our model maintains stability.

\subsubsection{Adiabatic Index}
The dynamical stability of the stellar model against an
infinitesimal radial adiabatic perturbation, which was introduced
by Chandrasekhar~\cite{chandrasekhar1964}, has also been tested in
our model. This stability condition was developed and used at
astrophysical level by several
authors~\cite{bardeen1966,knutsen,mak2013}.

The adiabatic index is defined by
\begin{equation}
\gamma=\left(\frac{\rho+p}{p}\right)\left(\frac{dp}{d\rho}\right).\label{eq34}
\end{equation}

For stable configuration $\gamma$ should be $>\frac{4}{3}$ within
the isotropic stellar system. However, we have analytically
calculated the value of the adiabatic index as $\gamma
=\frac{4}{3}$ which is the critical value of
$\frac{4}{3}$~\cite{chandrasekhar1964,Bondi1964,Wald1984}.

\subsection{Mass-Radius relation}
The mass function within the radius $r$ is given by
\begin{equation}
M(r)=\int_0^{r}4\pi r'^{2}\rho
dr'=\frac{4\pi}{\alpha}\left[-\frac{3kr^{(-\beta+1)}}{(-\beta+1)}
-\frac{3\sigma r}{\beta}+r\right].\label{eq35}
\end{equation}

The profile of the mass function has been depicted in Fig. 4,
which clearly shows that, for $r\rightarrow 0$, $M(r)\rightarrow
0$, implying the regularity of the mass function at the center.

\begin{figure}[h]
\centering
\includegraphics[width=0.4\textwidth]{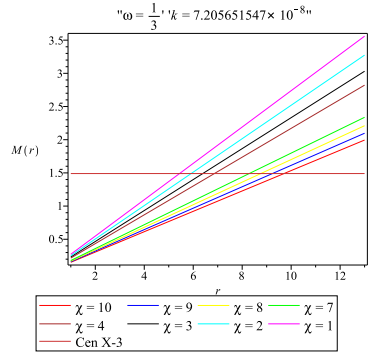}
\caption{Profile of the mass function $M(r)$ ($km$) is shown with
respect to the radial coordinate $r$ ($km$)}
\end{figure}

According to Buchdahl~\cite{buchdahl1959}, in the case of a static
spherically symmetric perfect fluid distribution the mass to
radius ratio $(\frac{2M}{r})$ should be~$\leq\frac{8}{9}$. Also
Mak et al.~\cite{mak2001} derived a more simplified expression for
the same ratio. In our present model, one can check that
Buchdahl's condition is satisfied (see Fig. 4).

\begin{figure}[h]
\centering
\includegraphics[width=0.4\textwidth]{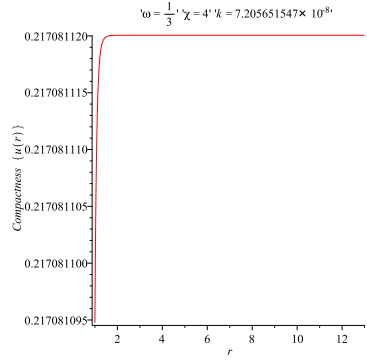}
\caption{Compactness $u(r)$ is plotted with respect to the radial
coordinate $r$ ($km$)}
\end{figure}

\subsection{Compactness and redshift}
The compactness of the star $u(r)$ is defined by
\begin{equation}
u(r)=\frac{M(r)}{r}=\frac{4\pi}{\alpha}\left[-\frac{3kr^{-\beta}}{(-\beta+1)}
-\frac{3\sigma}{\beta}+1\right].\label{eq36}
\end{equation}

The profile of the compactness of the star is depicted in Fig. 5.

\begin{figure}[h]
\centering
\includegraphics[width=0.4\textwidth]{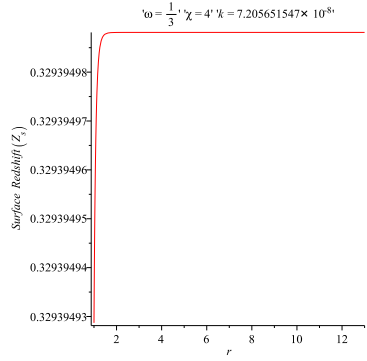}
\caption{Surface redshift ($Z_s$) is plotted with respect to the
radial coordinate $r$ ($km$)}
\end{figure}

The redshift function $Z_s$ is defined by
\begin{equation}
Z_s=(1-2u)^{-\frac{1}{2}} - 1 =
\left[1-\frac{8\pi}{\alpha}\left(-\frac{3kr^{-\beta}}{(-\beta+1)}
-\frac{3\sigma}{\beta}+1\right)\right]^{-\frac{1}{2}} -
1.\label{eq37}
\end{equation}

The profile of the redshift function of the star is depicted in
Fig. 6.

\section{A comparative study for physical validity of the model}
Based on the model under investigation let us carry out a
comparative study between the data of the model parameters with
that of the compact star candidates. This will provide the status
of the presented model as to whether it is valid for observed data
set within the allowed constraint. As we do not get the radius of
the star theoretically by putting $p =0$ at some radius,
therefore, all plots are drawn up to a highest calibrating point
of radius 13 km along the $r$-axis, which is sufficient to get
information as regards the nature of the compact star.

We have prepared Table 1 where the symbols are used as follows:
$D$ = observed radius, $M_{obs}$ = observed mass and $M_{pre}$ =
predicted mass. Here in calculation of $M_{pre}$ we have exploited
the observed radius $D$, the predicted radius being unable to be
determined in the present model as mentioned in the previous
paragraph. It is to note that we have drawn all figures assuming
$\chi =4$ only except the Fig. 4 for all $\chi$.

\begin{table*}[t]
\setlength\tabcolsep{2pt} \centering \caption{Comparative study of
the physical parameters for compact star and presented model for
$\omega=1/3$ and different values of $\chi $ and $k$.}
\begin{tabular}{@{}lllllllllll@{}}
\hline Compact & \hspace{0.8cm}$M_{obs}$ & \hspace{0.8cm}$D$ & $\chi$ & \hspace{1.2cm}$k$  & $M_{pre}$ & $\frac{M_{obs}}{D}$ & $\frac{M_{pre}}{D}$ & $Z_s (obs)$ & $Z_s (pre)$ & \\
Stars & \hspace{0.5cm}(in $M_\odot$) & \hspace{0.5cm}(in km) & & & \hspace{-0.2cm}(in $M_\odot$)& & & & &\\

\hline $4U 1608-52$ & 1.74 $\pm$ 0.14\cite{guver2010a} & 9.3 $\pm$ 1.0\cite{guver2010a}& 1 & $1.594868593\times 10^{-6}$ &  1.73 & 0.275968 & 0.273849 & 0.493929 & 0.486914 \\

\hline $Vela X-1 $ & 1.77 $\pm$ 0.08\cite{dey2013} & 9.56 $\pm$ 0.08\cite{dey2013} & 1 & $1.594868593\times 10^{-6}$ &  1.78 & 0.273091 & 0.274059 & 0.484428 & 0.487604 \\

\hline $4U 1820-30$ & 1.58 $\pm$ 0.06\cite{guver2010b} & 9.1 $\pm$ 0.4\cite{guver2010b}& 2 & $5.645562058\times10^{-7}$ &  1.55 & 0.256099 & 0.251890 & 0.431786 & 0.419590 \\

\hline $PSR J1903+327 $ & 1.667 $\pm$ 0.021\cite{dey2013} & 9.438 $\pm$ 0.03\cite{dey2013} & 2 & $5.645562058\times10^{-7}$ &  1.612 & 0.260521 & 0.251897 & 0.444945 & 0.419610 \\

\hline $Cen X-3 $ & 1.49 $\pm$ 0.08\cite{dey2013} & 9.178 $\pm$ 0.13\cite{dey2013} & 3 & $2.011363054\times10^{-7}$ &  1.45 & 0.239464 & 0.233199 & 0.385323 & 0.368962 \\

\hline $SMC X-4 $ & 1.29 $\pm$ 0.05\cite{dey2013} & 8.831 $\pm$ 0.09\cite{dey2013} & 4 & $7.205651547\times10^{-8}$ &  1.30 & 0.215468 & 0.217076 & 0.325621 & 0.329383 \\

\hline $PSR J1614-2230$ & 1.97 $\pm$ 0.04 \cite{Demorest2010}& 13 $\pm$ 2 \cite{Demorest2010} & 4 & $7.205651547\times10^{-8}$ & 1.91 & 0.223523 & 0.217085 & 0.344793 & 0.329404 \\

\hline $LMC X-4 $  & 1.04 $\pm$ 0.09 \cite{dey2013}& 8.301 $\pm$ 0.2\cite{dey2013} & 7 & $3.401707891\times10^{-9}$ &  1.01 & 0.184797 & 0.179810 & 0.259476 & 0.249629 \\

\hline $EXO 1785-248$ & 1.3 $\pm$ 0.2\cite{ozel2009} &11 $\pm$ 1\cite{ozel2009} & 7 &  $3.401707891\times10^{-9}$ & 1.34 & 0.174318 & 0.179809 & 0.239048 & 0.249627 \\

\hline $SAX J1808.4-3658$  & 0.9 $\pm$ 0.3\cite{dey2013} & 7.951 $\pm$ 1.0\cite{dey2013} & 8 & $1.238240496\times10^{-9}$ &  0.92 & 0.166960 & 0.170079 & 0.225284 & 0.231062 \\

\hline $4U 1538-52$  & 0.87 $\pm$ 0.07\cite{dey2013} & 7.866 $\pm$ 0.21\cite{dey2013} & 9 & $4.520512257\times10^{-10}$ &  0.86 & 0.163145 & 0.161340 & 0.218326 & 0.215075 \\

\hline $Her X-1$  & 0.85 $\pm$ 0.15\cite{dey2013} & 8.1 $\pm$ 0.41\cite{dey2013} & 10 & $1.654692258\times10^{-10}$ &  0.84 & 0.154790 & 0.153457 & 0.203492 & 0.201175 \\
\hline\label{tbl-1}
\end{tabular}
\end{table*}

Note that from the proposed model for $\chi=1-10$ (excluding 5 and
6 which do not provide physically interesting results) we have
found out the masses of the compact stars which, in general, are
closely equal to the observed values of most of the stars.
However, for some values of $\chi$ the model data seems not to
provide much significant results for some of the compact stars. It
is also interesting to note that in Fig. 4 we have the curve for
$\chi = 3$ and the straight line parallel to the $r$-axis for
$Cen~X-3$ total mass. So, the intersection of the two gives the
radius as a representaive one. However, the other curves for other
values of $\chi$ have no relation with the straight line parallel
to the $r$-axis. We also observe from Table 1 that for different
$\chi$ all the predicted values of Buchdahl's ratios fall within
the range of observed values of the Buchdahl ratios ($2M/R \leq
8/9 \sim 0.88$). On the other hand, the observed and predicted
values of the redshift are also very promising as is evident from
Table 1 for all the low mass compact stars under investigations.

\section{Discussions and conclusions}
As discussed in the introductory section, it is argued by
B\"{o}hmer et al.~\cite{Boehmer2011} that the $f(T)$ theory of
gravity with torsion scalar is more controllable than $f(R)$
theory of gravity with Ricci scalar because the field equations in
the former turn out to be the differential equations of second
order, whereas in the latter the field equations are in the form
of differential equations of fourth order and thus are difficult
to handle. On the other hand, the present work on
$f(R,\mathcal{T})$~\cite{harko2011} is based on another extension
of general relativity, which is associated to Ricci scalar $R$ and
the trace of the energy-momentum tensor $\mathcal{T}$.

At this juncture one may be curious to perform a comparison
between the results of our previous work~\cite{Das2015} on $f(T)$
gravity and the present work with $f(R,\mathcal{T})$ gravity.
However, we are at present very interested to present the model
behaviour of compact stars under the $f(R,\mathcal{T})$ theory of
gravity assuming the existence of CKV. In connection to the
features and hence validity of the model we have explored several
physical aspects based on our findings and all these have been
reflected to be very interesting advocacy in favor of physically
acceptance of the model. Let us now summarize some of these
important results as follows:

{\bf (i) Density and Pressure:} In the present investigation the
pressure $p$ and the density $\rho$ blow up as $r \rightarrow 0$
(Fig. 1). This clearly indicates that the core of the star is
highly compact and our model is valid for outside of the core. We
are unable to estimate the surface density as we do not find any
cut on the $r$-axis (i.e. the radius of the star) in the profile
of the pressure.

{\bf (ii) Energy conditions}: In our study we have found through
graphical representation that all the energy conditions, namely
NEC, WEC, SEC are satisfied within the prescribed isotropic fluid
distribution consisting normal matter (Fig. 2).

{\bf (iii) TOV equation}: The plot for the generalized TOV
equation reveals that static equilibrium has been attained by
three different forces viz. the gravitational force ($F_g$), the
hydrostatic force ($F_h$), and the additional force
($F_\varkappa$) (Fig. 3).

{\bf (iv) Stability of the model}: Following Herrera~\cite{herrera1992}
it has been observed that the squares of the
sound speed remains within the limit [0,1] admitting the condition
of causality and hence our model is potentially stable.

We have also studied dynamical stability of the stellar model
against the infinitesimal radial adiabatic perturbation where the
adiabatic index $\gamma$ has been calculated analytically as
$\frac{4}{3}$, which is the critical value for stable
configuration~\cite{chandrasekhar1964,Bondi1964,Wald1984}.

{\bf (v) Buchdahl condition}: The mass function within the radius
$r$ has been plotted in Fig. 4, which shows that, for
$r\rightarrow 0$, $M(r)\rightarrow 0$ implying the regularity of
the mass function at the center.

According to Buchdahl~\cite{buchdahl1959}, in the case of a static
spherically symmetric perfect fluid distribution the mass to
radius ratio $(\frac{2M}{r})$ should be~$\leq\frac{8}{9}$. In the
present model, we note that Buchdahl's condition is satisfied.

{\bf (vi) Compactness and redshift}: The profile of the compactness
of the star has been drawn in Fig. 5 whereas the redshift function
$Z_s$ of the star has been depicted in Fig. 6. The features as
revealed from these figures are physically reasonable.

As one of the major concluding remarks we would like to highlight
one special observation that in the present model the profile of
the density and the pressure (Fig. 1) reveals that both the
density and the pressure suffer from central singularity.
Therefore we are unable to make any exact comment on the core of
the star, though Figs. 1 and 5 also indicate a high compactness of
the core. On the other hand, according to the profile of the mass
function (Fig. 4) it maintains the regularity at the center.

Another interesting point can be observed from the assumed data
for $\omega=1/3$ which represents an equation of state (EOS) for
radiation. However, in the present investigation we have tried to
explore other values of the EOS parameter $\omega$ but those do
not work well. This seems to indicate that our model suits better
for radiating compact stars. In favour of this unique result one
can go through some supporting
literature~\cite{Demorest2010,LA2001,SM2001,Govender2003,SH2006,ARR2016}.
But this also immediately raises the problem of the energy
conservation in the model. As is well known, in the
$f(R,\mathcal{T})$ gravity theory the energy-momentum tensor is
not conserved [see Eq. (3)]. This means we may have two probable
alternatives: (i) either we must fully investigate and present the
energy ``conservation" equations for the present model and discuss
their possible interpretation as describing radiation emission
from the star, (ii) otherwise by maintaining the problem of
conservation we have to give up the claim for radiating compact
stars in our study assuming that the case for $\omega=1/3$ is just
a coincidence out of other several choices of $\omega$ . These
intriguing issues may be taken into consideration in a future
project.

\section*{Acknowledgments} FR and SR are thankful to the
Inter-University Centre for Astronomy and Astrophysics (IUCAA),
India for providing Visiting Associateship under which a part of
this work was carried out. SR is thankful to the authority of The
Institute of Mathematical Sciences (IMSc), Chennai, India for
providing all types of working facility and hospitality under the
Associateship scheme. FR is also grateful to DST-SERB and
DST-PURSE, Government of India for financial support. We all are
very thankful to the anonymous referee for several useful
suggestions which have enabled us to revise the manuscript
substantially.

\end{document}